\documentstyle[12pt,fleqn]{article}
\begin{document}
\title{
The Density Matrix Renormalization Group for Fermion Systems
}
\author{ R.M.\ Noack and S.R.\ White \\
Department of Physics\\
University of California, Irvine, CA 92717 \\ \\
D.J. Scalapino \\
Department of Physics\\
University of California, Santa Barbara, CA 93106
}

\maketitle
\begin{abstract}
We discuss techniques of the density matrix renormalization group and their
application to interacting fermion systems in more than one dimension.
We show numerical results for equal--time spin--spin and
singlet pair field correlation functions, as well as the spin gap for
the Hubbard model on two chains.
The system is a gapped spin liquid at half--filling and shows weak algebraic
$d$-wave--like pair field correlations away from half--filling.
\end{abstract}

\section{Introduction}

The numerical renormalization group was
developed by Wilson \cite{wilson1} and used by him to
solve the one impurity Kondo problem.
The technique was subsequently applied to a number of quantum lattice
systems \cite{earlyrg,lee1} such as the Hubbard and Heisenberg models, but
with little success.
A suggestion by Wilson \cite{wilson2} to investigate why the technique
fails
for the simplest quantum lattice system, the one-dimensional electron
gas, led to the development of a number of new techniques to
overcome the difficulties of the numerical RG for this simple system
\cite{whitenoack}.
White \cite{whitedmrg} was able to generalize one of these
techniques to interacting systems, applying it successfully to one
dimensional quantum spin systems.
This technique has come to be known as the density matrix
renormalization group (DMRG).

This paper describes our our current efforts to apply the DMRG to
fermion systems in more than one dimension, and in particular to the
Hubbard model.
So far, we have successfully applied the method to the Hubbard model
on one and two chains \cite{twochains}.
Here we discuss the details of the methods we have developed for the
two--chain Hubbard model and show results for equal--time pair field
and spin--spin correlation functions and for the spin gap for the
half--filled and doped systems on lattices of up to $2\times 32$ sites.

At half--filling, both pair field and spin--spin correlations
decay exponentially, with the spin correlations having a longer
correlation length.
There is a spin gap present at half filling which gets smaller as the
system is doped, but persists down to band fillings of
$\langle n \rangle =0.75$.
For the doped system, the largest pair field correlations are ones in
which a spin singlet pair is formed on adjacent sites on different
chains.
The pair field symmetry is $d$--wave--like in that the pair field wave
function has opposite sign along and between the chains.
The form of the decay of the pair field correlations for the doped
system is algebraic with a form close to that of the
noninteracting system, which decays as $\ell^{-2}$.

\section{The Density Matrix Renormalization Group}

The goal of the procedures discussed here is to find the
properties of the low-lying states of a quantum system on a particular
finite lattice.
One way to do this would be to diagonalize the Hamiltonian matrix
using a sparse matrix diagonalization
method such as the Lanczos technique.
However, for interacting quantum lattice systems, the number of states
grows exponentially with the size of the lattice.
Since exact diagonalization techniques must keep track of all the
states, the maximum possible lattice sizes for interacting Hamiltonians is
severely limited.
It is therefore desirable to develop a procedure in which the Hilbert space
of the Hamiltonian can be truncated in a controlled way so that only
states that are important in making up the low-lying states of the
system are included in a diagonalization.
The DMRG provides a procedure for building up such a representation of
the Hamiltonian matrix, which is then diagonalized to provide the
properties of the low-lying states of the finite system.

The strategy of the DMRG is to build up a portion
of the system (called the system block)
using a real--space blocking procedure and then truncate the basis of its
Hamiltonian after each blocking.
In this way, the size of the Hilbert space is kept manageable as the
system block is built up.
The key idea is the method of truncating the Hilbert space of the system block
in a controlled way.
This is done by forming the reduced density matrix for the system
block, given an eigenstate of the entire lattice.
Let us first examine this procedure.

\subsection{The Density Matrix Projection }

Consider a complete system (the ``universe''), divided into
two parts, the
``system'', labeled by coordinate $i$, and the ``environment''
\cite{liang}, labeled by coordinate $j$.
If we knew the exact state $\psi_{ij}$ of the universe,
(assuming the universe is in a pure state) the
prescription for finding the state of the system block would be to form
the reduced density matrix of the system as part of the universe,
\begin{equation}
\rho_{ii'} = \sum_j \psi_{ij} \psi^*_{i'j}.
\label{denmateqn}
\end{equation}
The state of the system block is then given by a linear combination of
the eigenstates of the density matrix with weight given by the
eigenvalues.
It is shown in Ref. \cite{whitedmrg} that the optimal
reduced basis set for the system block is given by
the eigenstates of the density matrix with the largest weights.
The sum of the density matrix weights of the discarded states gives the
magnitude of the truncation error.

\section{Algorithms}

The density matrix projection procedure gives us a way of truncating
the basis set of the matrix for the system block in a controlled way
as degrees of freedom are added to the system.
The projection procedure of the previous section assumes that the
wavefunction $\psi_{ij}$ of the system is known.
Of course, finding $\psi_{ij}$ is the goal of the DMRG procedure, so
effective algorithms must iteratively improve
approximations to $\psi_{ij}$.
We will first discuss the algorithms for one-dimensional systems, as
developed in Ref.\ \cite{whitedmrg}.

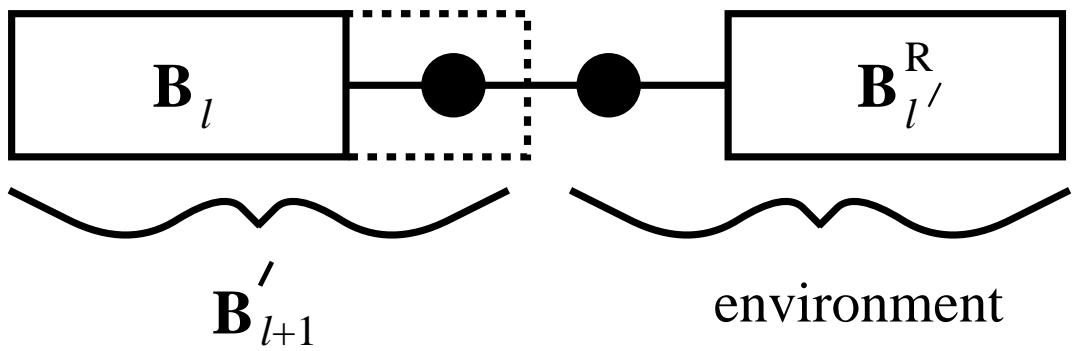
\begin{figure}
\caption{
The superblock configuration for the one--dimensional algorithms.
The solid circles represent single sites treated exactly and the boxes
represent approximate Hamiltonians representing $\ell$ and $\ell'$ sites.
}
\label{figsuper1d}
\end{figure}

In order to perform the density matrix projection procedure, we form
the Hamiltonian for a
``superblock'' which is an approximation to the universe of the previous
section.
In this case, the superblock will describe a one-dimensional lattice of
$L$ sites, with, for example, a Heisenberg or Hubbard Hamiltonian.
The superblock configuration used for the
one--dimensional algorithms developed in Ref. \cite{whitedmrg} is
shown in Fig.\ \ref{figsuper1d}.
The superblock is formed from an approximate Hamiltonian for the system block
containing $\ell$ sites (labeled by $B_\ell$), the Hamiltonians for
two single sites which can be treated exactly, represented by solid
circles, and an approximate Hamiltonian for the rightmost $\ell'$
sites, labeled by $B^R_{\ell'}$.
Thus, the superblock contains $L = \ell + \ell' + 2$ sites.
The algorithm proceeds as follows:
\begin{description}
\item{1.} The superblock Hamiltonian is diagonalized using a Lanczos
or similar exact diagonalization technique to find a particular
target eigenstate $\psi_{ij}$.
\item{2.} The reduced density matrix is formed for the system block
$B'_{\ell+1}$ using Eq.\ (\ref{denmateqn}).
\item{3.} The density matrix is diagonalized using a dense matrix
diagonalization.
\item{4.} The Hamitonian for $B'_{\ell+1}$ is transformed to a
truncated basis formed by the $m$ highest weighted eigenstates of the
density matrix.
\item{5.} This approximate Hamiltonian, labeled by $B_{\ell+1}$ is
used as a starting point for the next iteration, starting with step 1.
\end{description}
Initially we choose $\ell$ to be small enough (a single
site, for example) so that the
Hamiltonian for $B_\ell$ can be treated exactly.
The system block then grows by a single site at each iteration, but
the dimension of its Hilbert space remains $m$.
A single site only is added to $B_\ell$ in order to minimize the size
of the superblock Hamiltonian, whose dimension will be $n^2mm'$ where
$n$ is the number of states per site, and $m'$ is the size of the
basis for $B^R_{\ell'}$.

\subsection{The Infinite System Procedure}

The method we use to choose $B^R_{\ell'}$  at each step divides DMRG
algorithms into two classes, the infinite system procedure and the finite
system procedure.
In the infinite system procedure, $B^R_{\ell'}$ is chosen to be the spatial
reflection of $B_\ell$ so that $\ell=\ell'$.
This means that the size $L$ of the superblock grows by two sites
at each iteration.
The procedure can be iterated until the energy, calculated in the
superblock diagonalization, converges.

The advantage of the infinite system procedure is that calculated
quantities scale to their infinite system values.
In this sense, this procedure is in the spirit of the original
real--space renormalization group.
The disadvantages of the infinite system procedure are that for a
given system size, it is less accurate than the finite system
procedure, and that it cannot easily be generalized to
two--dimensional systems.
For a two-dimensional system, if a single site is added to the
system block at each step, an environment block of the proper
geometry cannot in general be formed from the reflected system block.

\subsection{The Finite System Procedure}

In the finite system procedure, the superblock is formed so that it
describes the same finite lattice at each iteration.
In other words, the block $B^R_{\ell'}$ is chosen so that
$L=\ell + \ell' + 2$ remains fixed.
We can do this if we repeat the procedure (which we call a sweep
through the lattice) in which the system block is
built up from $\ell = 1$ to $\ell = L-3$ more than once.
After one sweep, the system block can be built up from the other side
of the lattice, and the stored set of
system blocks from that sweep can be used as environment blocks for
the next sweep.
The procedure is analogous to zipping a
zipper back and forth once through the lattice,
where the location of the zipper is the
location of the single site added to $B_\ell$.
The sweeps can be repeated  until the energy or some other quantity of
interest converges.
In practice, we have found that it only takes a few sweeps through the
lattice to achieve convergence to within truncation error for a given
$m$.
The power of this procedure lies in the iterative improvement of the
environment block.

On the initial sweep of the finite system procedure, the
environment blocks are undefined.
For one dimensional systems, however, one can build up the superblock
size using the infinite system procedure and use reflections of the stored
blocks $B_\ell$ for $B^R_{\ell'}$ on the initial sweep.

There are a number of advantages to the finite system procedure.
First, since the environment blocks are iteratively improved with each
sweep through the lattice, the finite system procedure gives
much more accurate results for a particular lattice size than the
infinite system procedure, although the infinite system procedure can
give results that are closer to the thermodynamic limit.
It might be possible to combine the two procedures in a hybrid
algorithm to get more accurate results for a given $m$ in the
thermodynamic limit.

Second, since the environment block no longer must be a reflection of the
system block, it is possible to study lattices that are no longer
reflection symmetric.
This is useful, for example, in studying systems with impurities or
disorder.

Third, in the finite system procedure, the target state of the
superblock is the same at each iteration, with unchanging quantum
numbers, unlike in the infinite system procedure.
For the one-dimensional Heisenberg model calculations described in
Ref. \cite{whitedmrg}, the states are labeled only by the $z$ component of
the total spin, $S_z$, so it is easy to find a state with the
appropriate quantum number for different lattice sizes.
For fermion systems such as the Hubbard model, however, $N_\uparrow$
and $N_\downarrow$, the number of spin up and spin down fermions, are
good quantum numbers.
Since $N_\downarrow$ and $N_\uparrow$ must be integers, it is
impossible to choose them so that the overall occupation stays
constant on all different lattice sizes, except at half--filling.
The best one can do is to target one or more states closest to the
proper density, and this leads to reduced accuracy for
non--half--filled systems.

Fourth, it is much easier to extend the finite system procedure to
lattices of more than one dimension.

\subsection{Extension to Higher Dimensions}

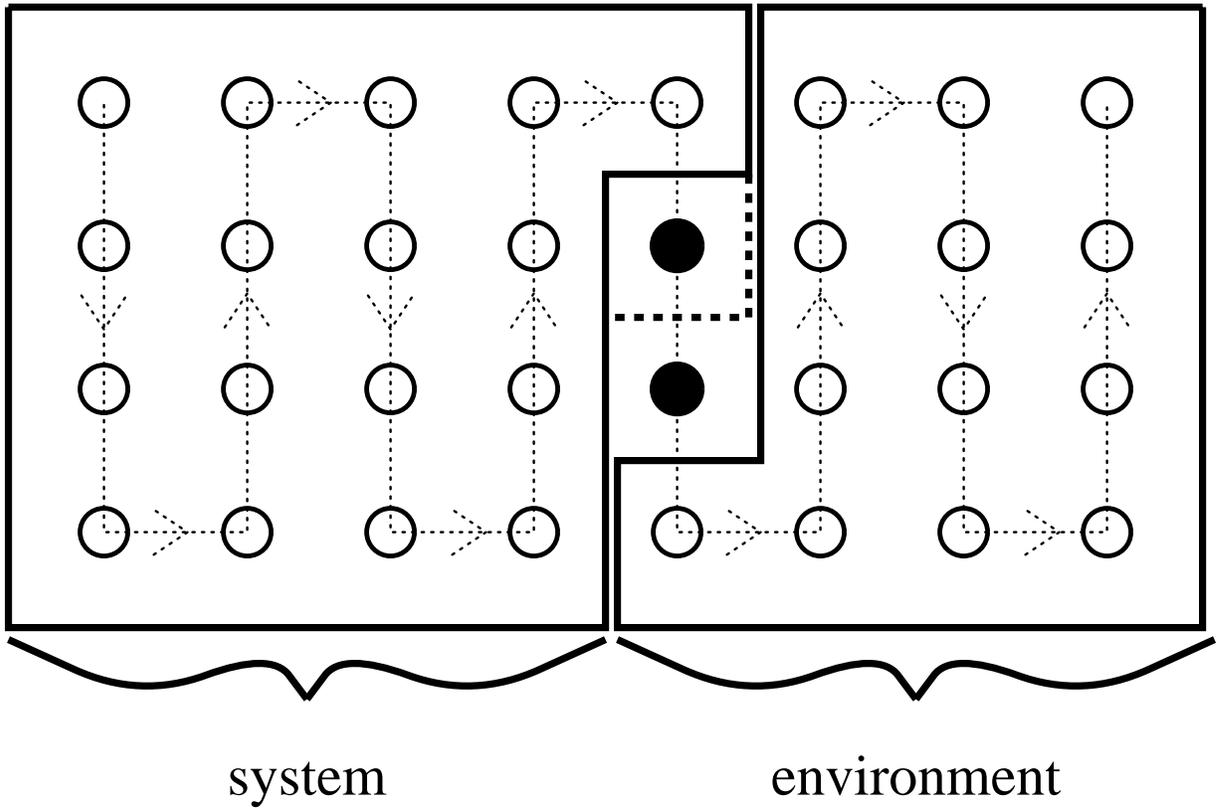
\begin{figure}
\caption{
The superblock configuration for the two--dimensional algorithm.
The order in which sites are added to the system block on a series of
iterations is given by the dotted line, and the site added to the
approximate system block Hamiltonian is outline by the dashed line.
}
\label{figsuper2d}
\end{figure}

One way to extend these algorithms to more than one dimension would be
to replace the single sites added between the blocks with a row of
sites.
However, the extra degrees of freedom added to the system at each
real--space blocking would make size of the superblock Hilbert space
prohibitively large.
Therefore, the two--dimensional algorithms we have developed still
involve adding single sites at a time to the system block.
This can be done by adding sites in a connected one-dimensional path
through the two--dimensional lattice, i.e. by folding the
one-dimensional zipper into two dimensions.
A typical superblock configuration for the two-dimensional algorithm
is shown in Fig.\ \ref{figsuper2d}.
The site added to the system block is enclosed by a dashed line and
the dotted line shows the order in which sites are added to the system
block for a sweep.
One can see that it is not possible to reflect the system block into
an environment block of the proper geometry at every iteration, so the
finite system algorithm must used.
The two--dimensional procedure differs from the one--dimensional
finite size procedure only in that there are additional connections
between the system and environment blocks along the boundary.


For one--dimensional lattices, we use the infinite system procedure
to build up the superblock to the proper size on the first sweep through the
lattice.
Since this can no longer be done for higher dimensional lattices, we
must formulate a procedure for the initial sweep through the lattice.
The simplest procedure is to use an empty environment block on
the first sweep.
One can diagonalize the Hamiltonian for the system block and keep the $m$
states of lowest energy.
This procedure is equivalent to Wilson's original numerical
renormalization group procedure, and is not very accurate even for
the one--dimensional single electron on a lattice, as shown in
Ref. \cite{whitenoack}.
In addition, for fermion systems, one must adjust the chemical
potential $\mu$ so
that states with the proper $N_\uparrow$ and $N_\downarrow$ quantum
numbers have the lowest energy.
The procedure is quite sensitive to these adjustments.
Thus, this initialization technique thus tends to be inaccurate and hard to
use for fermion systems.

Liang \cite{liang2} has tried two other techniques for the initial
sweep.
In the first, he performs an initial infinite system sweep for a
one-dimensional lattice, then turns on the additional couplings needed
to make the lattice two dimensional on subsequent finite system sweeps.
In the second, he uses as the environment block an approximate
Hamiltonian for a one--dimensional system of the size of the row
length.
Both of these procedures depend on representing portions of
two-dimensional states by one-dimensional states, and thus
give poor representations of the superblock initially.

The technique which we find works best for the initial sweep is a
hybrid procedure in which the finite system procedure for a smaller
lattice size is repeated for a few iterations, until the system block is
big enough so that its reflection can be used for the environment
block of a superblock that is a row larger.
Thus, the superblock is extended a row at a time.
Initially, the first row can be built up with a one-dimensional
infinite system procedure.
This procedure minimizes problems with target states with
inappropriate quantum numbers and provides a reasonable representation
for two--dimensional states.

We have found that the accuracy of the initial sweep is not critical
as long as the first set of environment blocks has a set of states
with appropriate quantum numbers.
In most cases, a few sweeps of the finite system procedure will
improve the environment blocks sufficiently so that the procedure will
converge.

\subsection{Performance Considerations}

The number of states needed to maintain a certain truncation error
in the density matrix projection procedure depends strongly on
the number of operators connecting the two parts of the system.
Best accuracy is obtained when the number of connections between the
system and environment blocks is minimized.
Therefore, we study systems with open rather than periodic or
antiperiodic boundary conditions.
Also, we find that the number of states $m$ needed to maintain a given
accuracy depends strongly on the width and weakly on the length of the
system.

Just how rapidly the truncation error increases with the width of the
system is not clear in general.
Liang \cite{liang2} studied the error in the energy as a function
of width for a gas of noninteracting spinless fermions and found that
the number of states needed to maintain a given accuracy grew
exponentially with the width of the system.
In an interacting system such as the Hubbard model, the detailed
structure of the energy spectrum seems to be important.
For example, in the two chain Hubbard
model at half--filling, where there is a spin and pairing gap,
the truncation error for a given $m$ is much
smaller than away from half--filling, where the spin gap is reduced
and the gap to pairing excitations is no longer present.
For multiple Hubbard or Heisenberg chains, the presence or absence of
a gap in the spin spectrum depends on whether the number of chains is
even or odd\cite{twospinchains}, so the truncation error for a given $m$
depends on the number of chains in a complicated way.
Also, increasing the strength of on-site interactions
can reduce the truncation error.
The Hubbard model DMRG is most accurate for large $U$ and least
accurate for $U=0$.

For systems of more than one dimension, it is therefore important to
be able to keep as many states $m$ per block as possible.
We have been able to improve the performance of the algorithm in a
number of ways.
One way of doing this is to minimize the size of the superblock
Hilbert space, whose dimension is $n^2mm'$.
For fermion systems, one can reduce the number of states per site $n$
from four to two by treating the spin degree of freedom on the same
footing as a spatial coordinate.
A site for a particular spatial coordinate and spin can have an
occupancy of zero or one fermion.
While this makes the path through the lattice (which now has an added
dimension) somewhat more
complicated, we have found that by adding these ``half--sites'' instead
of full spatial sites on the last few sweeps through the lattice
we can increase the accuracy by increasing $m$.
We have also found that $m'$ can be made smaller than $m$ without
losing much accuracy in the truncation\cite{whitedmrg}.
Since the representation the approximate block Hamiltonians is poor on
the first few sweeps through the lattice, making $m$ large initially
does not improve the representation very much.
Therefore, the most efficient procedure is to increase $m$ after every
sweep through the lattice, so that $m'$ is $m$ from the previous sweep.

We have made a major effort to write the code in an efficient way in
C++.
We store only the nonzero parts of operators that link states with
particular quantum numbers.
These matrices are dense in general because the basis transformation
at each step mixes matrix elements.
This representation minimizes memory usage and makes it possible to
optimize highly the multiplication of a vector by the Hamiltonian, the
basic step needed for the Lanczos diagonalization.
However, the resulting data structures are complicated and are
variable in size, so that it has been useful to take advantage of
the object--oriented data structures and dynamic memory allocation
available in C++.
The code is currently limited more by memory usage than by computer
time, although we minimize memory usage by writing to disk all
operators not needed for a particular superblock diagonalization step.
The current version of the code can handle $m=400$ or more
whereas the original Fortran code used in Ref.\ \cite{whitedmrg}
for the computationally less demanding Heisenberg spin problem could keep
at most $m=200$.
We have found that $m \approx 400$ is necessary
in order to obtain accurate results for the two--chain Hubbard model away
from half--filling.

\section{Results for the two--chain Hubbard model}

The two--chain Hubbard model is described by the Hamiltonian
\begin{eqnarray}
\begin{array}{cl}
H= &
- t_y \sum_{ i, \lambda \sigma}
( c^{\dagger}_{i,\lambda \sigma}c_{i+1,\lambda \sigma} +
 c^{\dagger}_{i+1,\lambda\sigma}c_{i,\lambda \sigma}  ) \\
&- t_x \sum_{i, \sigma}
( c^{\dagger}_{i,1 \sigma}c_{i,2 \sigma } +
 c^{\dagger}_{i, 2\sigma}c_{i, 1 \sigma} )
+ U\sum_{ i, \lambda}
n_{i,\lambda \uparrow}n_{i,\lambda \downarrow } . \\
\end{array}
\end{eqnarray}
We think of the lattice as beinging a ladder aligned with the $y$ axis
so that $c^{\dagger}_{i, \lambda \sigma}$ creates an electron of spin
$\sigma$ at rung $j$ and side
$\lambda=1$ (left) or $2$ (right),
the hopping along a chain is $t_y$, the hopping between chains on a
rung is $t_x$, and $U$ is the on--site Coulomb repulsion.
This system is thought to be relevant to a number of anisotropic
two--dimensional systems, including
$({\rm VO})_2 {\rm P}_2 {\rm O}_7$ \cite{johnston} and
${\rm Sr}_2 {\rm Cu}_4{\rm O}_6$ \cite{takano,rice1}, which have
weakly coupled ladder--like structures arranged in planes.
Here we will concentrate on a parameter regime relevant to
the latter class of substances: $U/t_y=8$, and $t_x=t_y$.
We will explore the phase diagram as a function of band filling as the
half--filled system is doped with holes.

At half--filling,
the Hubbard model maps to the Heisenberg model in the large $U/t_y$
limit.
Therefore, the
dominant correlations should be antiferromagnetic spin correlations.
However, it is known that in the Heisenberg model on two chains
\cite{dagotto,barnes,strong}, there
is spin gap leading to an exponential decay of the spin correlation
function.
The origin of the spin gap is easy to understand in the limit of
strong coupling across the rungs.
In this case, the only only interaction will be an
antiferromagnetic coupling between the two spins on a rung.
This two spin system forms a spin singlet state and a higher energy
triplet state with an energy separation of the Heisenberg coupling $J$.
Away from half--filling, it is not clear what correlations
dominate the behavior.
Some authors \cite{rice2,tsunetsugu} have predicted that singlet
superconductivity with a partial d-wave symmetry should be the
dominant order.

In order to resolve these issues, we have calculated equal time
spin--spin and pair field correlation functions
$S_{\lambda\lambda '}(i,j)=\langle M^z_{i,\lambda} M^z_{j,\lambda '}\rangle$,
$D_{xx}(i,j)= \langle \Delta_{x i} \Delta^\dagger_{x j} \rangle$,
and
$D_{yx}(i,j) = \langle \Delta_{y i} \Delta^\dagger_{x j} \rangle$
with
\begin{eqnarray}
\begin{array}{cl}
M^z_{i,\lambda} & = n_{i,\lambda \uparrow} - n_{i,\lambda \downarrow}\\
\Delta^\dagger_{x i} & = c^\dagger_{i,1 \uparrow} c^\dagger_{i,2 \downarrow}
- c^\dagger_{i,1 \downarrow} c^\dagger_{i,2 \uparrow}  \\
\Delta^\dagger_{y i} & =
 c^\dagger_{i+1,2 \uparrow} c^\dagger_{i,2 \downarrow}
- c^\dagger_{i+1,2 \downarrow} c^\dagger_{i,2 \uparrow}. \\
\end{array}
\end{eqnarray}
Here $S_{11}(i,j)$ and $S_{12}(i,j)$ measure the spin-spin
correlations along a chain and between the chains respectively,  and
$D_{xx}(i,j)$ measures the singlet pair field correlations in which a
singlet pair is added at rung $j$ and removed at rung $i$.
In addition, $D_{yx} (i,j)$ measures the pair field correlations in
which a singlet pair is added to rung $j$ and removed from the
right--hand chain between rungs $i$ and $i+1$.
The relative phase of the pair wave function across the $i$th rung to
along one chain from $i$ to $i+1$ is given by comparing the phase of
$D_{xx}(i,j)$ to $D_{yx}(i,j)$.
This turns out to be negative, corresponding to the mean field result
obtained in Ref. \cite{rice2}.
However, the non-interacting $U=0$ result at a filling
$\langle n \rangle = 0.875$ is also negative.

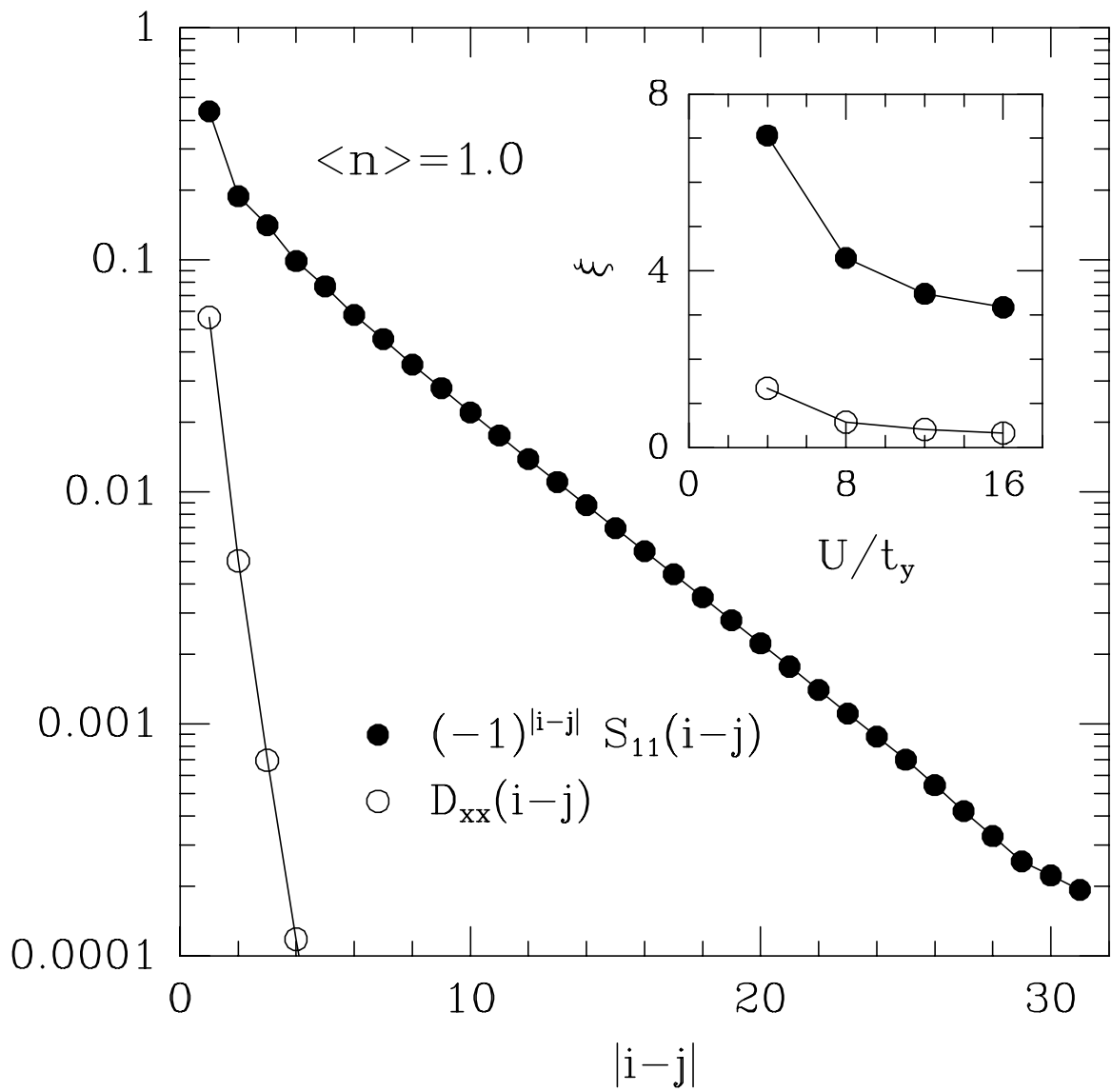
\begin{figure}
\caption{
Semilog plot of the spin--spin correlation function $S_{11}(i-j)$ and
the pair field correlation function $D_{xx}(i-j)$ at half--filling and
$U/t_y=8$.
The insert shows the correlation lengths in units of the lattice
spacing obtained from similar plots for various $U/t_y$ values.
}
\label{figsemilog}
\end{figure}

Fig.\ \ref{figsemilog} shows the logarithm of the antiferromagnetic spin--spin
correlation function
\hbox{$(-1)^{|i-j|}S_{11}(i-j)$} and the cross--chain pairing correlation
function $D_{xx}(i-j)$.
Both the correlation functions decay
exponentially with $|i-j|$, but the pair field correlations decay much more
rapidly.
The correlation length, calculated from the slope of the lines in the
semilog plot, is plotted as a function of $U/t_y$ in the inset.
The spin--spin correlation length decreases as $U$ is increased,
saturating at a value near 3 lattice spacing for large $U$.
We have calculated the spin--spin correlation length for the
isotropic two chain Heisenberg model using the DMRG
\cite{twospinchains} and find a value of 3.19 lattice spacings,
consistent with the large $U$ limiting value.
The pair field correlations decay with a correlation length of the
order of a lattice spacing and are thus negligible at half--filling.

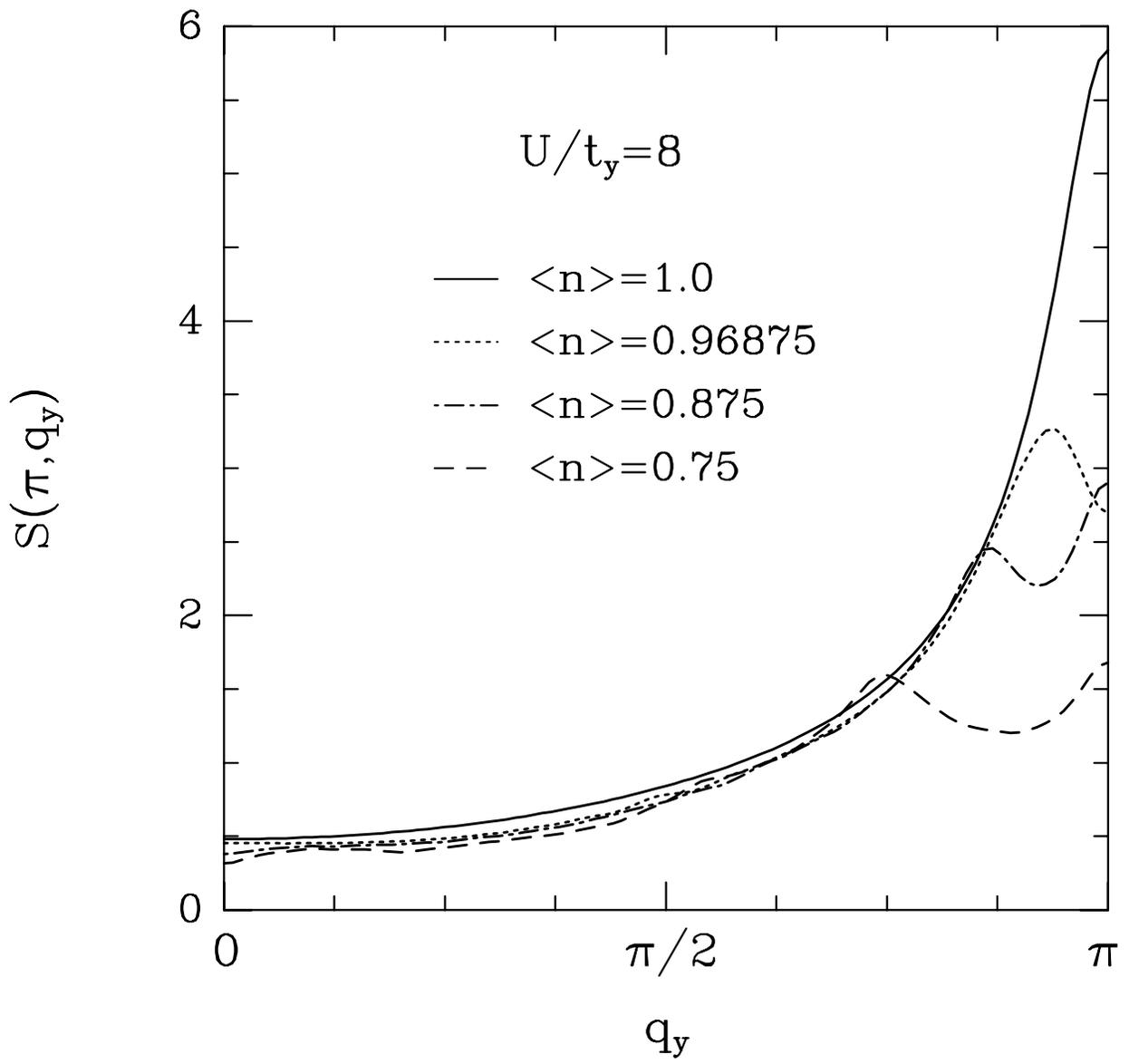
\begin{figure}
\caption{
The fourier transform $S(\pi,q_y)$
of the spin--spin correlation function $S_{\lambda \lambda'}(\ell)$.
Here $t_x=t_y$ and $U/t_y=8$ and the calculations were made on a
$2 \times 32$ lattice.
}
\label{figSq}
\end{figure}

In order to determine the behavior of the spin correlations as the
system is doped below half-filling, we have calculated the magnetic
structure factor $S(q_x,q_y)$ by taking the fourier transform of
$S_{\lambda\lambda '}(i,j)$.
Since the lattice is long in the $y$ direction
and the spin--spin
correlation function decays exponentially with $ | i - j | $,
one can take a continuous fourier transform in the $y$ direction
without introducing much error.
Since there are two chains, $q_x$ can take on the values $0$ and
$\pi$.
Only the $S(\pi,q_y)$ branch is interesting,
because the correlations are always antiferromagnetic across the rungs.
This function is plotted in Fig.\ \ref{figSq} for the fillings,
$\langle n \rangle = 1.0, 0.9875, 0.875, 0.75$, corresponding to
doping 0, 2, 8, and 16 holes into the half--filled $2\times 32$ lattice.
As the system is doped away from half-filling, $S(\pi,q_y)$ peaks at
a wavevector $q_y = \langle n \rangle \pi$.
The residual peak at $q_y=\pi$ present for
$\langle n \rangle = 0.875$ and $\langle n \rangle = 0.75$ is
present only for even numbers of hole pairs and thus probably
disappears in the thermodynamic limit.
Therefore, we see that the spin--spin correlations develop
incommensurate structure as the system is doped away from
half--filling.

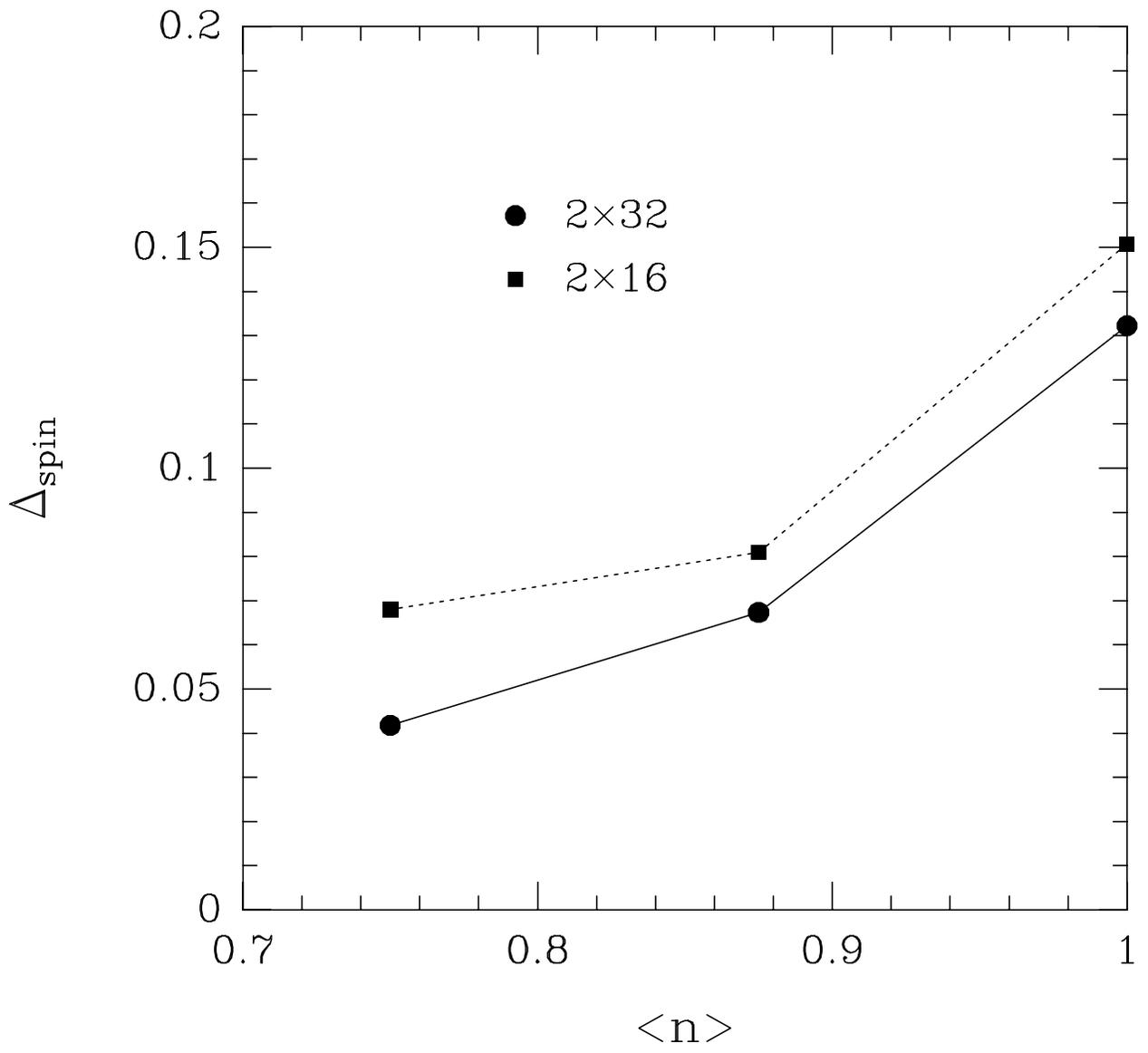
\begin{figure}
\caption{
The spin gap $\Delta_{\rm spin}$ plotted as a function of band filling
$\langle n \rangle$ on a $2 \times 32$ lattice for $U/t_y = 8$ and
$t_x=t_y$.
}
\label{figspingap}
\end{figure}

One can calculate the spin gap directly, by calculating the difference
in energies between the ground state, which has total spin $S=0$, and the
lowest lying $S=1$ state.
We calculate the ground state energy for $N_\uparrow$ spin up electrons and
$N_\downarrow$ spin down electrons,
$E_0(N_\uparrow,N_\downarrow)$.
The spin gap for a system with $N_\uparrow=N_\downarrow=N$ electrons
is then given by $\Delta_{\rm spin} = E_0(N+1,N-1) - E_0(N,N)$.
The spin gap plotted as a function of filling is shown in
Fig.\ \ref{figspingap}.
It is largest at half--filling and becomes smaller as the system
is doped with holes and seems to be present at least down to fillings of
$\langle n \rangle = 0.75$.
We show the spin gap for $2 \times 16$ and $2 \times 32$ lattices to
show the size of the finite size effects and argue that they are small
enough that the gap is present in the thermodynamic limit for two
chains.

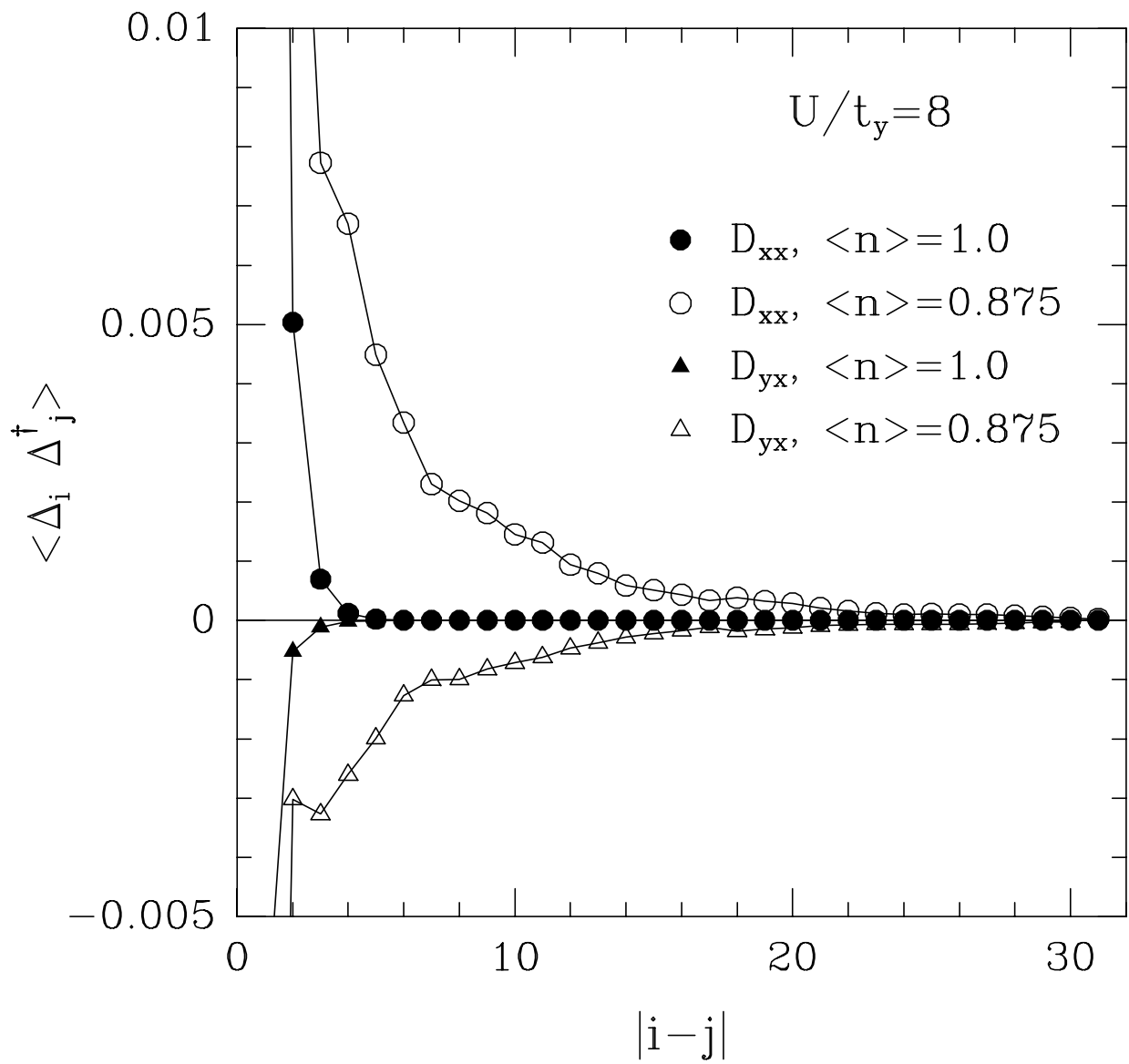
\begin{figure}
\caption{
The pair field correlation functions $D_{xx}(i-j)$ and $D_{yx}(i-j)$
versus $|i-j|$ on a $2\times 32$
lattice with $U/t_y = 8$ and $t_x=t_y$.
}
\label{figDij}
\end{figure}

We now turn to the behavior of the pair field correlations as the
system is doped away from half--filling.
We have seen that the pairing correlations with cross--chain symmetry
decay exponentially in the half--filled system.
This is true for all symmetries of the pair field wavefunction.
Fig.\ \ref{figDij} shows the pair field correlations $D_{xx}(i-j)$ and
$D_{yx}(i-j)$ plotted as a function of $|i-j|$ for
$\langle n \rangle = 1.0$ and $\langle n \rangle = 0.875$.
One can see that $D_{xx}(i-j)$ and $D_{yx}(i-j)$ have opposite signs,
as one would expect for $d$-wave like symmetry, at both fillings and
are significantly enhanced for the doped system.

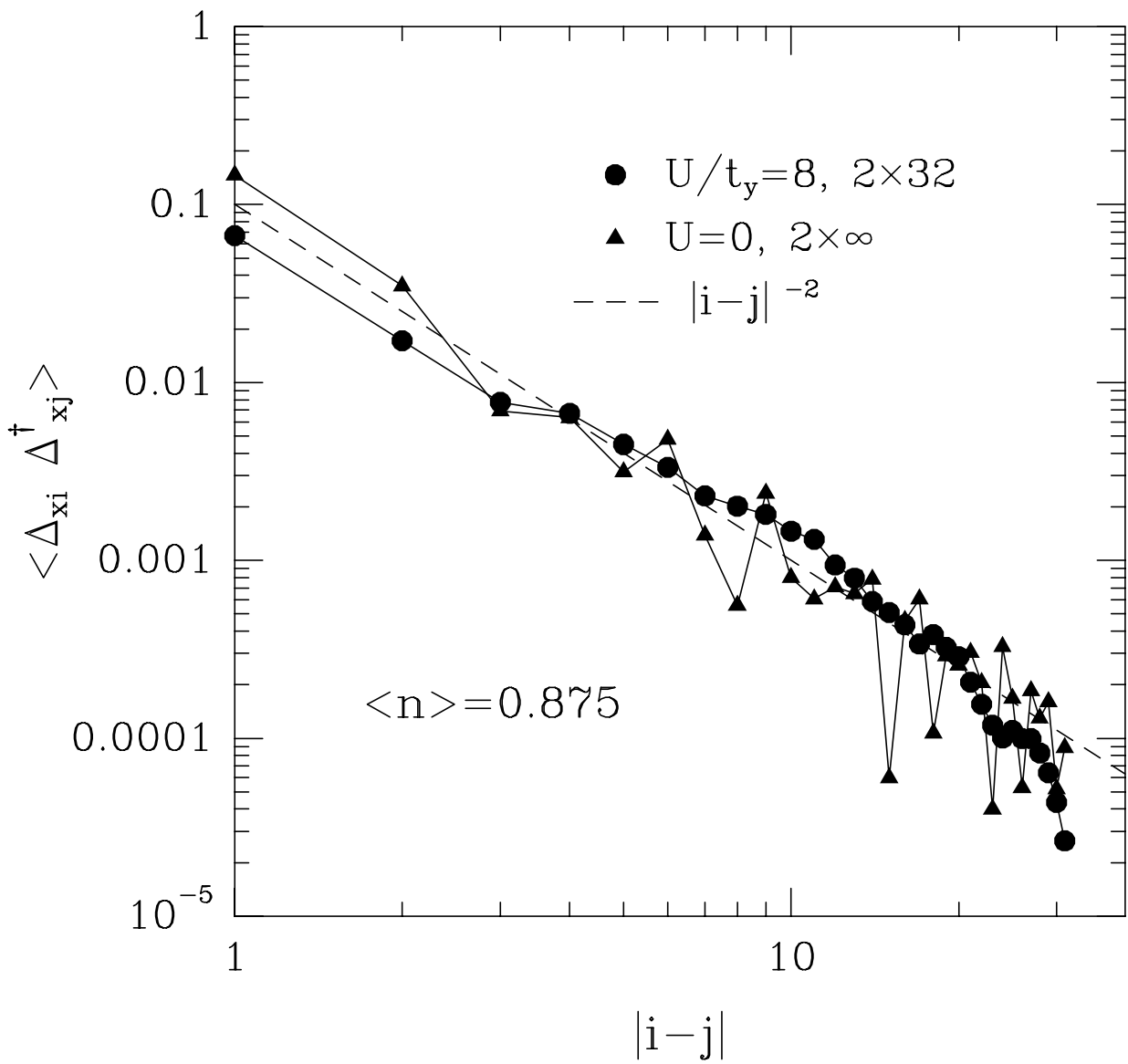
\begin{figure}
\caption{
Log-log plot of the rung--rung single pair field correlation function
$D_{xx}(i-j)$
versus $| i-j |$ for a $2 \times 32$ cluster with $U/t_y=8$ and an infinite
two--chain system with $U=0$.
In both cases $\langle n \rangle = 0.875$.
The dashed line shows $|i-j|^{-2}$.
}
\label{figllDx}
\end{figure}

In order to determine the strength of the pairing correlations,
one must consider their $\ell$--dependence at large distances.
For a quasi--one--dimensional system, we expect that any pairing correlation
will at best decay as a power of $\ell$ and can in some cases
decay exponentially, as we have seen for the half-filled system.
For two chains, one can compare with the
the non-interacting $U=0$ ladder, for which
\begin{equation}
D_{xx}(\ell) = ( 1/2 \pi \ell )^2
\left[ 2 - \cos ( 2 k_f(0)\ell) - \cos (2 k_f(\pi) \ell) \right].
\label{uzpair}
\end{equation}
Here $k_f(0) = \cos^{-1} (t_x+\mu)/2$ and $k_f(\pi) = \cos^{-1} (t_x-\mu)/2$
are the Fermi wave vectors corresponding to the bonding and
antibonding bands of the two coupled chains with $\mu$ the chemical
potential.
The pair correlations, $D_{xx}(\ell)$ are shown in Fig.\
\ref{figllDx}, plotted on a log--log scale.
The correlations of the interacting system decay approximately as
$\ell^{-2}$ and do not seem to be significantly enhanced over those of
the non--interacting system, as given by Eq.\ (\ref{uzpair}).

\section{Conclusion}

We have discussed techniques we have developed to apply the density
matrix renormalization group to Fermion systems in more than one
dimension.
In particular, we have been able to obtain accurate results for energy
gaps and equal--time correlation functions for the Hubbard model on
two coupled chains.

The two--chain Hubbard model is a gapped spin liquid at half--filling.
Both spin--spin and pair field correlations decay
exponentially, with the spin--spin correlations having the longest
correlation length.
As the system is doped with holes, the spin--spin
correlations become incommensurate at a wave vector proportional to
the filling and the spin gap becomes smaller, but persists in the
thermodynamic limit.
The pairing correlations are enhanced with a $d$-wave--like symmetry
and decay algebraically with an exponent close to that of the
non--interacting, $U=0$ system.

\section*{Acknowledgements}
The authors thank N. Bulut, T.M. Rice, A. Sandvik, M. Vekic, E.
Grannan, and R.T. Scalettar for useful discussions.
R.M.N. and S.R.W. acknowledge support from the Office of Naval
Research under grant No. N00014-91-J-1143 and D.J.S. acknowledges support
from the National Science Foundation under grant DMR92--25027.
The numerical calculations reported here were performed at
the San Diego Supercomputer Center.

\newpage

\newcommand{\prb}{{\it Phys. Rev. B }}
\newcommand{\prl}{{\it Phys. Rev. Lett. }}

\newpage

\vfill


\begin{thebibliography}{99}
\bibitem{wilson1} K.G. Wilson, {\it Rev. Mod. Phys.}  47, 773 (1975).
\bibitem{earlyrg} J.W. Bray and S.T. Chui, {\it Phys. Rev. }{\bf B}19,
4876 (1979);
S.T. Chui and J.W. Bray, {\it Phys. Rev. }{\bf B}18, 2426 (1978);
J.E. Hirsch,{\it Phys. Rev. }{\bf B}22, 5259 (1980);
C. Dasgupta and P. Pfeuty, {\it J. Phys. }{\bf C}14, 717 (1981).
\bibitem{lee1} P.A. Lee, {\it Phys. Rev. Lett. } {\bf 42 }, 1492 (1979).
\bibitem{wilson2} K.G. Wilson, in an informal seminar.
\bibitem{whitenoack} S.R.\ White and R.M.\ Noack, \prl {\bf 68} 3487, (1992).
\bibitem{whitedmrg} S.R. White, \prl {\bf 69}, 2863 (1992),
\prb {\bf 48}, 10345 (1993).
\bibitem{twochains} R.M.\ Noack, S.R.\ White, and D.J.\ Scalapino (to be
published).
\bibitem{liang} The term ``environment'' block is due to Shoudan
Liang.
\bibitem{liang2} S.\ Liang (to be published).
\bibitem{twospinchains} S.R.\ White, R.M.\ Noack, and D.J.\ Scalapino
(to be published).
\bibitem{johnston} D.C. Johnston {\it et al.}, \prb {\bf 35}, 219 (1987).
\bibitem{takano} M. Takano, Z. Hiroi, M. Azuma, and Y. Takeda, Jap. J.
of Appl. Phys. Series {\bf 7}, 3 (1992).
\bibitem{rice1} T.M. Rice, S. Gopalan, and M. Sigrist, Europhys. Lett.
{\bf 23}, 445 (1993).
\bibitem{dagotto} E.\ Dagotto, J.\ Riera, and D.J.\ Scalapino, \prb
{\bf 45}, 5744 (1992).
\bibitem{barnes} T.\ Barnes et al., \prb {\bf 47}, 3196 (1993).
\bibitem{strong} S.P.\ Strong, and A.J.\ Millis, \prl {\bf 69}, 2419 (1992).
\bibitem{rice2} M.\ Sigrist, T.M.\ Rice, and F.C.\ Zhang (to be published);
Sudha Gopalan, T.M.\ Rice, and M.\ Sigrist (to be published).
\bibitem{tsunetsugu} H. Tsunetsugu, M. Troyer, and T.M.\ Rice (to be
published).

\end{thebibliography}
\end{document}